\documentclass[english]{article}
\usepackage{mathptmx}
\usepackage[T1]{fontenc}
\usepackage[latin9]{inputenc}
\usepackage[a4paper]{geometry}
\usepackage{color}
\usepackage{babel}
\usepackage{url}
\usepackage{amsmath}
\usepackage{amssymb}
\usepackage[unicode=true]
 {hyperref}

\makeatletter
\newcommand{\lyxaddress}[1]{
\par {\raggedright #1
\vspace{1.4em}
\noindent\par}
}

\makeatother

\begin{document}
\global\long\def\ckmffitter{{\sc CKM4fitter}}
\newcommand{\slfrac}[2]{\left.#1\right/#2}

\title{Status of the Fourth Generation\\*\large{\emph{A Brief Summary of B3SM-III Workshop in Four Parts}}}

\author{S.\,A. \c{C}etin$^{1}$, G.\,W.-S. Hou$^{2,3}$, V.\,E. \"Ozcan$^{4}$,
A.\,N. Rozanov$^{5}$, S. Sultansoy$^{6,7}$}

\maketitle
%\begin{doublespace}

\lyxaddress{\begin{center}
$^{1}$ Physics Department, Do\u{g}u\c{s} University, Istanbul, Turkey\\*
$^{2}$ Department of Physics, National Taiwan University, Taipei, Taiwan\\*
$^{3}$ National Center for Theoretical Sciences, National Taiwan University, Taipei, Taiwan\\*
$^{4}$ Department of Physics, Bo\u{g}azi\c{c}i University, Istanbul, Turkey\\*
$^{5}$ CPPM (IN2P3-CNRS-Universite de Mediterranee), Marseille, France\\*
$^{6}$ Physics Department, TOBB ETU, Ankara, Turkey\\*
$^{7}$ Institute of Physics, Academy of Sciences, Baku, Azerbaijan
\par\end{center}}
%\end{doublespace}

\begin{abstract}
This summary of the 3rd ``\emph{Beyond the 3 Generation Standard Model}''
workshop presents the following four statements (and their implications)
for the ongoing and future searches of a fourth generation:
 1) The enhancement of the Higgs gluon-gluon production cross-section
  times branching fraction for many of the search channels studied at
  the Large Hadron Collider (LHC) is not a flat factor of 9;
 2) Electroweak precision data allows for not only a fourth generation,
  even more generations are allowed;
 3) Consideration of mixing significantly changes the conclusions about
  the interpretation of experimental constraints, and even a fully-degenerate
  fourth generation becomes allowed;
 4) The features that make a fourth generation of fermions attractive are
  still valid even under the light of initial LHC results.
\end{abstract}

\section*{Introduction}

It is well known that the number of fundamental fermion families (generations)
is not fixed within the Standard Model (SM).
In recent years, with the discovery of neutrino oscillations, and
reevaluations and reconsideration of electroweak (EW) precision data,
it became evident that a fourth generation of fermions is
not ruled out experimentally.
Furthermore, such an extra generation was identified to have
theoretically attractive features, and could help address
some of the fundamental open questions of nature.

These arguments have been summarized briefly in$~$\cite{b3sm1},
which is itself a summary of the first ``\emph{Beyond the 3-generation SM in
the LHC era}'' workshop held at CERN on 4-5 September 2008.
Here we try to provide a brief summary of some of the key topics
discussed in the third iteration of this thematic series, carrying the banner
``\emph{Third Workshop on Beyond 3 Generation Standard Model --- Under
the light of the initial LHC results}'' (B3SM-III),
which was held at Bo\u{g}azi\c{c}i University in Istanbul on 23-25 October 2011.
The full agenda and further details of the presentations can be found in~\cite{indico}.

Recent Tevatron and LHC results on the direct search of the fourth
generation quarks, and indirect limits from perceived enhancement
of $gg\rightarrow H$ cross-section, significantly reduce the allowed
phase-space of chiral fourth generation theories. Therefore, more
accurate formulations of the models and assumptions under which the
limits are derived are needed to arrive at the final conclusion on
the existence or absence of the fourth generation.

\subsection*{Enhancement of the Higgs discovery channels}

Given that the gluon-gluon fusion process, i.e. the dominant production
channel of the Higgs boson at the Large Hadron Collider (LHC), contains
a quark loop, the addition of two new heavy quarks beyond the top
quark would approximately triple the amplitude, and thus increase the
production cross-section by a factor of 9. This simple description
is valid unless the Higgs boson is quite heavy ($m_{H}\gtrsim2m_{t}$),
in which case the contribution from the top to the production amplitude
turns imaginary, and thus the enhancement of the amplitude-squared
decreases to be around 5 ($2^{2}+1^{2}$ instead of $3^{2}$). This behaviour
is illustrated in~\cite{cetinHiggs} for a number of scenarios.

However, the quantity that is relevant to the Higgs searches at the
LHC is not simply $\sigma(gg\rightarrow H)$, but the product
$\kappa_{\, X}=\sigma(pp\rightarrow H)\times {\cal B}(H\rightarrow X\bar X)$.
This product can have an enhancement factor that is significantly
lower than $5\sim9$ (even less than to 1 in some cases):
\begin{itemize}
\item The $ZZ$, $WW$, and\footnote{
 In the $H\rightarrow\tau^{+}\tau^{-}$ analysis channel, the selection
 criteria often preferentially keep events produced through the vector
 boson fusion process, so effective enhancement would be even smaller
 than what is discussed here.%
 }
$\tau^{+}\tau^{-}$ final states:
The Higgs boson can decay into the new fermions if they are lighter
than half the Higgs mass. While this has relatively little effect
for a very heavy Higgs boson, $\kappa\,_{W,Z,\tau}^{\mathrm{SM4}}$ can
be smaller than $\kappa\,_{W,Z,\tau}^{\mathrm{SM3}}$ if the
mass of the fourth generation neutral lepton is close to the
current experimental bounds\footnote{
 Exact limit extracted from just the Z-lineshape is $m_{\nu_{\tau}'}>46.7$ GeV
 at 95\% CL~\cite{lepNu4mass}.%
 }
($50\sim60$ GeV) and the Higgs boson is light ($m_{H}\lesssim150$ GeV)~\cite{sashaHiggs,salehHiggs}.
\item The $\gamma\gamma$ final state:
Unlike the gluon-gluon fusion process, the loop diagram contains
not only heavy fermions (including the heavy charged lepton),
but also the weak boson as well, and destructive interference cancels out
most of the enhancement in production, leading to $\kappa\,_{\gamma}^{\mathrm{SM4}}\approx\kappa\,_{\gamma}^{\mathrm{SM3}}$.
\end{itemize}
Therefore, care needs to be taken in the interpretation of the recent ATLAS and CMS results
on the Higgs search$~$\cite{LHCHiggs}. Certainly, a conclusion
like ``the idea of a sequential fourth generation of quarks and leptons
is in serious trouble''$~$\cite{peskin} is quite premature.
Non-observation of the enhanced $gg\rightarrow H$ signal may as well be
the signature of the Higgs absence (or its very high mass as in
warped$~$\cite{frankWarpedHiggs} and other scenarios),
rather than the signature of the absence of the fourth generation.
Final conclusion should be reserved for an update with higher luminosity.
Moreover, the results from the search for the Higgs will not be fully
conclusive unless they are complemented with direct searches for the
fourth-family fermions themselves, as the presence of a light neutral lepton
or an extended scalar sector (as in$~$\cite{extendedScalar}) can
still resolve the tension between a sequential fourth generation and
measurements of a SM-like Higgs.
Finally, in the case of a vector-like new generation,\footnote{
 New vector-like quarks might offer a number of attractive properties.
 For an overview, see$~$\cite{gustavoVectors}.
 }
where the quarks have mass terms that do not originate from EW symmetry breaking
(such as the model presented in$~$\cite{vectorLike}), there
is no enhancement of $gg\rightarrow H$ production,
and the Higgs constraints would be very different.

\subsection*{Generations beyond the Fourth}

Since the fourth generation was itself incorrectly considered to be
inconsistent with the EW precision data,
generations beyond the fourth were considered to be outright impossible.
However, as the fourth generation was shown to have %finite
sufficiently wide allowed parameter space,
possibilities of considering more generations have recently been revisited.
In$~$\cite{rozanov_extragen}, a fit to the EW data using the
{\sc Leptop} code$~$\cite{leptop} for the example scenario of
$m_{t'}=m_{b'}=300$ GeV, $m_{\tau'}=200$ GeV,\footnote{
 We use the notation of \cite{b3sm1}, with primed symbols
 for third-generation fermions representing their fourth generation
 counterparts.%
 }
showed that, while unfavored, two extra generations with Dirac-type
neutrinos had not been ruled out yet.
Analysis of the Peskin-Takeuchi oblique parameters $S$, $T$
using the {\sc Opucem} code$~$\cite{opucem} confirms this finding,
even for $m_{t'}=m_{b'}=500$ GeV or higher.
Furthermore, it is found that if the neutrinos of the extra generations
are of Majorana nature, it is possible to accommodate even more than
5 generations with $S$, $T$ values within 1$\sigma$ of the LEP
EWWG measurements$~$\cite{opucem5gen}.

If a SM-like Higgs boson is assumed to exist, the current absence
of any Higgs signal from the LHC obviously undermines the possibility
of quarks from extra generations.\footnote{In the case of infinitely
heavy new fermions, the enhancement factor for the production of a 120\,GeV
Higgs boson via gluon-gluon fusion is 8.5, 24 and 47 in SM with four, five
and six generations respectively~\cite{arikHiggs}.} Therefore 
the fact that such extra
generations are not excluded by the EW data could be interpreted as
a pointer to extending only the lepton sector in a SM-like Higgs scenario.
It is worth noting that there is theoretical motivation for such extra
generations. For example, in$~$\cite{aparici}, it is shown that
with 5 generations and two right-handed neutrinos it might be possible
to generate all present data on the observed light neutrino mass hierarchy.

\subsection*{Consequences of Flavor Mixings}

Implications of dropping the unitarity requirement from the $3\times3$
PMNS matrix were pointed out in$~$\cite{pmns4}.
Precision of the measured values of the Fermi constant $G_{F}$
and CKM parameter $|V_{ud}|$ are significantly reduced.
The $2\sigma$ range for the $4\times4$ PMNS parameter
$U_{e4}$ is $0.021<|U_{e4}|<0.089$, and the p-value for $|U_{e4}|=0$
(as would be expected in SM with 3 generations) is only 2.6\%.
Such large mixings has itself some interesting implications:
in the absence of any other physics, experimental constraints from
neutrinoless double beta decay would imply Dirac or pseudo-Dirac
(nearly degenerate mass eigenstates obtained when the
Majorana mass is much smaller than the Dirac mass) nature\footnote{
 Further constraints (which apply not only to fourth generation) have
 recently been extracted on the grounds that lepton number violation
 due to the interactions involving new Majorana neutrinos can wash
 out considerably any GUT scale generated or otherwise preexisting
 baryon asymmetry of the universe$~$\cite{neutrino_BAU}.%
}
for the fourth generation neutrinos$~$\cite{neutrinoMixings}.

In the quark sector, the importance of considering quark mixings has
recently been discussed in$~$\cite{lenzPaper}. As the quark mixings
increase the $T$ parameter without changing the $S$ parameter, it
becomes possible to accommodate even a fully-degenerate fourth family,
contrary to the widely quoted sentiment in$~$\cite{pdg}.\footnote{Almost
degenerate fourth family quarks and charged lepton are predicted by the
Flavor Democracy Hypothesis~(see~\cite{flavdemocracy} and references
therein.)}

In light of all these findings, it is clear that a fully consistent picture
cannot be drawn unless the flavor and EW data are considered together.
Furthermore, a consequence of considering fermion mixings is that new
physics may not enter only through gauge boson self-energies,
therefore considerations of only the oblique parameters are not enough$~$\cite{non-oblique},
and a new generic approach for computing the EW corrections
is being developed to be included in the global fitter program, \ckmffitter$~$\cite{lenzTalk}.

Preliminary results from the first global fit using the \ckmffitter~
indicate that the currently available data favors values around 0.1-0.2
for the $4\times4$ CKM parameters $|V_{cb'}|$ and $|V_{tb'}|$$~$\cite{lenzTalk}.\footnote{
 These large values seem to contradict those found in~\cite{HouMa},
 probably because of not yet considering $Z\to b\bar b$ constraint.
 }
This finding has important consequences for the direct searches at the LHC,
as the currently most stringent mass constraints on the fourth-generation
quarks assume either ${\cal B}(t'\rightarrow Wb) = 100\%$ or
${\cal B}(b'\rightarrow Wt)=100\%$$~$\cite{CMShq,ATLAShq}.
%,
%whereas the limits that do not make assumptions on the mixings are
%much weaker$~$\cite{ATLAShq}.
An independent (but simpler) analysis by the {\sc Opucem} code yields
very similar results for the mixings as those of the \ckmffitter,
and also indicate a tendency for decreasing
${\cal B}\sim\slfrac{|V_{tb'}|^{2}}(|V_{tb'}|^{2}+|V_{cb'}|^{2})$ as the
quark masses increase$~$\cite{opucemMixings}.

Finally, irrespective of what is inferred from the EW and CKM data,
it is clear that the mass limits obtained from direct searches
for new quarks and leptons should always be given as a function of
the mixing angles. In particular in the case of the small mixing angles
of the fourth generation quarks with the other generations the detection
efficiency can be reduced due to displaced vertices or quasi-stable
fourth generation hadrons. This will require specialized search strategy
and interpretations.

\subsection*{Fourth Generation, Open Questions and New Models}

In$~$\cite{b3sm1}, it was highlighted that a fourth generation could
address or provide clues about a number of open problems (baryon asymmetry
of the universe, Higgs naturalness, fermion mass hierarchy, dark matter).
Most of these arguments were reviewed in this workshop and found to
be still valid and interesting in the light of the recent LHC data$~$\cite{HouAndSoni}.
Amongst these, the topic that received the most attention was the role
that the fourth generation quarks could play in EW symmetry breaking
through some strong dynamics,\footnote{
 Dynamical EW symmetry breaking without the presence of some strong
 dynamics has been most recently been considered in$~$\cite{vaqueraDEWSB},
 but the recent experimental results seem to exclude such a scenario.%
 }
since direct searches have pushed the quark mass lower limits
close to 500$\,$GeV~\cite{CMShq}, not far from the so-called unitary bound.

Surprisingly, little is known about a strong Yukawa theory. Different
groups have recently been exploring different aspects of the strong
regime through various approaches. For example, following the second
B3SM workshop in January 2010~\cite{b3sm2}, a collaboration was formed to
study the Higgs-Yukawa theory on the lattice in the regime of large bare
Yukawa coupling; their recent progress is summarized in$~$\cite{lattice}.
Schwinger-Dyson approach is taken in$~$\cite{pqSDE}, which allows
the computation of the self-energy of the fourth generation fermions
and subsequently their condensates; the results obtained are consistent
with the interpretation of the perturbative evolution of the Yukawa
couplings$~$\cite{pqRGE}. Estimates for the phenomenology of Yukawa-bound
$Q\bar{Q}$ states are made in$~$\cite{houBound}, using a relativistic
Bethe-Salpeter equation approach to motivate the Yukawa binding energy
before the unitarity bound sets in. This work also concludes
that the current search for pair-produced fourth generation quarks
can continue beyond the unitarity bound, as the leading $gg\rightarrow Q\bar{Q}$
fusion is found to not exhibit significant resonance phenomena.

A fourth generation is an attractive tool for model building. The
explanations it can provide for some of the open questions of physics
is applicable irrespective of whether there is beyond the SM (BSM)
physics or not. Many of its indirect consequences, such as the enhancements
in neutron EDM$~$\cite{neutronEDM}, \mbox{$\mu$-e} conversions~\cite{mueConv},
$B_{s}\rightarrow\mu\mu$ branching fraction$~$\cite{bs2mumu}, etc.,
have also been computed and found to be within current experimental bounds,
while $B_{s}\rightarrow\mu\mu$ and mixing-dependent CPV in $B_s \to J/\psi\phi$ 
might still exhibit indications for the fourth generation~\cite{bs2mumu},
despite the recent ``setback" from LHCb~\cite{lhcb}.
%Therefore a 
A number of models combine fourth generation with BSM ideas
to address further experimental hints from the colliders.
For example, a new phenomenological model involving strongly coupled
fourth generation quarks has been proposed in~\cite{soniTopAFB}
to address top-pair forward-backward asymmetry observed at the Tevatron
experiments$~$\cite{tevTopAFB}. It is interesting to note that the
experimental signatures of the fourth generation in new models can
be quite unusual and might require significant rethinking of conventional
searches at the LHC$~$\cite{flavdemocracy,soniT2th}.

\section*{Conclusion}

\begin{center}
\emph{``The rumors of my death have been greatly exaggerated, again!''
- Fourth Generation}
\par\end{center}

We are approaching very fast the moment when the LHC searches will
discover the fourth generation or exclude it in the TeV mass scales.
The latter requires careful searches in multiple channels and full
phase space of masses and mixing angles. From an experimental point
of view, these searches are needed irrespective of the indirect constraints,
which have been misinterpreted multiple times in the past. If no sign
of the fourth generation is found at the LHC, the firm experimental
proof of only 3 generations would be the fundamental result which
will require satisfactory theoretical explanation beyond the Standard
Model.

\subsection*{Acknowledgments}

We thank all the participants for their contributions, and for the
lively discussions that their presentations stimulated. The local
organization committee acknowledges the financial support from the
Bo\u{g}azi\c{c}i University Foundation and Bo\u{g}azi\c{c}i Physics
Department.


\begin{thebibliography}{10}
\bibitem{ATLAShq}
A review of the current ATLAS results in the searches
for fourth-generation quarks was presented by
S. Grancagnolo, %``ATLAS results on $4{}^{\mathrm{th}}$
%generation and combined Higgs searches'', %\url{https://indico.cern.ch/contributionDisplay.py?sessionId=0&contribId=4&confId=150154}
Ref.~\cite{indico}.
\end{thebibliography}
\end{document}